\documentclass[aip,reprint,jap]{revtex4-1}

\usepackage{graphicx} \usepackage{amsmath, amssymb, bm}
\usepackage{dcolumn} \usepackage{color}
\usepackage{fix-cm}
\usepackage{times}
\usepackage{multirow}
\usepackage{sidecap}

\usepackage{marginnote}
\usepackage{ulem} 

\newcommand{\rzcm}{cm$^{-1}$}
\renewcommand{\emph}[1]{\textit{#1}}

\begin{document}
\title{Optical phonon modes, static and high frequency dielectric constants, and effective electron mass parameter in cubic In$_2$O$_3$}
\author{Megan Stokey}
\email{mstokey@huskers.unl.edu}
\homepage{http://ellipsometry.unl.edu}
\affiliation{Department of Electrical and Computer Engineering, University of Nebraska-Lincoln, Lincoln, NE 68588, USA}
\author{Rafa\l{} Korlacki}
\affiliation{Department of Electrical and Computer Engineering, University of Nebraska-Lincoln, Lincoln, NE 68588, USA}
\author{Sean Knight}
\affiliation{Department of Electrical and Computer Engineering, University of Nebraska-Lincoln, Lincoln, NE 68588, USA}
\affiliation{Terahertz Materials Analysis Center and Center for III-N Technology, C3NiT -- Janz\`{e}n, Department of Physics, Chemistry and Biology (IFM), Link\"{o}ping University, 58183 Link\"{o}ping, Sweden}
\author{Alexander Ruder}
\affiliation{Department of Electrical and Computer Engineering, University of Nebraska-Lincoln, Lincoln, NE 68588, USA}
\author{Matthew Hilfiker}
\affiliation{Department of Electrical and Computer Engineering, University of Nebraska-Lincoln, Lincoln, NE 68588, USA}
\author{Zbigniew Galazka}
\affiliation{Leibniz-Institut f\"{u}r Kristallz\"{u}chtung, 12489 Berlin, Germany}
\author{Klaus Irmscher}
\affiliation{Leibniz-Institut f\"{u}r Kristallz\"{u}chtung, 12489 Berlin, Germany}
\author{Yuxuan Zhang}
\affiliation{Department of Electrical and Computer Engineering; and Department of Materials Science and Engineering, The Ohio State University, Columbus, OH 43210 USA}
\author{Hongping Zhao}
\affiliation{Department of Electrical and Computer Engineering; and Department of Materials Science and Engineering, The Ohio State University, Columbus, OH 43210 USA}
\author{Vanya Darakchieva}
\affiliation{Terahertz Materials Analysis Center and Center for III-N Technology, C3NiT -- Janz\`{e}n, Department of Physics, Chemistry and Biology (IFM), Link\"{o}ping University, 58183 Link\"{o}ping, Sweden}
\author{Mathias Schubert}
\affiliation{Department of Electrical and Computer Engineering, University of Nebraska-Lincoln, Lincoln, NE 68588, USA}
\affiliation{Terahertz Materials Analysis Center and Center for III-N Technology, C3NiT -- Janz\`{e}n, Department of Physics, Chemistry and Biology (IFM), Link\"{o}ping University, 58183 Link\"{o}ping, Sweden}
\affiliation{Leibniz Institut f\"{u}r Polymerforschung e.V., 01069 Dresden, Germany}

\date{\today}

\begin{abstract}
A complete set of all optical phonon modes predicted by symmetry for bixbyite structure indium oxide is reported here from a combination of far-infrared and infrared spectroscopic ellipsometry, as well as first principle calculations. Dielectric function spectra measured on high quality, marginally electrically conductive melt grown single bulk crystals are obtained on a wavelength-by-wavelength (a.k.a. point-by-point) basis and by numerical reduction of a subtle free charge carrier Drude model contribution. A four-parameter semi-quantum model is applied to determine all sixteen pairs of infrared-active transverse and longitudinal optical phonon modes, including the high-frequency dielectric constant, $\varepsilon_{\infty}=4.05\pm 0.05$. The Lyddane-Sachs-Teller relation then gives access to the static dielectric constant, $\varepsilon_{\mathrm{DC}}=10.55\pm 0.07$. All experimental results are in excellent agreement with our density functional theory calculations and with previously reported values, where existent. We also perform optical Hall effect measurements and determine for the unintentionally doped $n$-type sample a free electron density of $n=(2.81 \pm 0.01)\times 10^{17}$~cm$^{-3}$, mobility of $\mu=(112 \pm 3)$~cm$^{2}$/(Vs), and an effective mass parameter of $(0.208\pm0.006)m_e$. Density and mobility parameters compare very well with results of electrical Hall effect measurements. Our effective mass parameter, which is measured independently of any other experimental technique, represents the bottom curvature of the $\Gamma$ point in In$_2$O$_3$ in agreement with previous extrapolations. We use terahertz spectroscopic ellipsometry to measure the quasi-static response of In$_2$O$_3$, and our model validates the static dielectric constant obtained from the Lyddane-Sachs-Teller relation. 
\end{abstract}

\maketitle
\section{Introduction}

Indium oxide (In$_2$O$_3$) is a sesquioxide of continued interest as a part of the broader transparent semiconducting oxide family. In$_2$O$_3$ crystallizes in a stable phase with bixbyite structure and cubic crystal symmetry. Similarly to other transparent semiconducting oxides, In$_2$O$_3$ is classified as a wide bandgap (2.9~eV) material.\cite{King09} When doped with tin, In$_2$O$_3$ shows high electrical conductivity.\cite{Tahar98, Shigesato92, Bierwagen14} These properties lend In$_2$O$_3$ to applications in transparent electrodes, \cite{Minami05,LUNGENSCHMIED07, DINIZ2011, KIM2009} gas sensing,\cite{Bartic2007, TISCHNER2008, KANNAN2010} nanowire technology,\cite{Nguyen2004} high-voltage transistors,\cite{Nayak2013, Han2011, Kim2016} Schottky diodes,\cite{Sheu1998,Zhang2015Nature} and ultra-violet light emitting devices,\cite{Wei2010, Gao_2011, Ouacha_2017, Chen_2011} for example.


In$_2$O$_3$ with high structural quality has been produced by thin film and bulk growth methods. Most existing works study thin film samples and as a result, most of the applications are designed around thin film samples as well.\cite{NAGATA2019} Bierwagen and Speck reported high quality thin films grown by plasma-assisted molecular beam epitaxy (PAMBE);\cite{Bierwagen14,Bierwagen_2015} Yang~\textit{et al.} reported similarly high quality thin films grown by metal organic vapor phase epitaxy;\cite{YANG20084054} and Karim, Feng, and Zhao reported high quality thin films grown by low pressure chemical vapor deposition,\cite{Karim8b00483} for example. Recently, Galazka \emph{et al.} developed a high quality single crystal bulk growth method, and thereby provided substrates with large surface areas for epitaxy.\cite{GALAZKA2013} The structural quality of epitaxial layers grown, e.g., by PAMBE, is typically higher than for polycrystalline films and lower than or comparable to bulk crystals. With increasing availability of bulk samples, the possibilities for homoepitaxial and heteroepitaxial growth are becoming abundant.\cite{Sadofev2012, Liu2019}


In$_2$O$_3$ is commonly $n$ type and unintentionally doped (UID). High quality thin films and bulk crystals, UID and doped with tin, were investigated recently by Feneberg~\textit{et al.}\cite{Feneberg2016} The free electron density varied from $3.7\times 10^{17}-1.2\times10^{21}$~cm$^{-3}$ and can be well controlled. Electrically insulating high quality single crystalline In$_2$O$_3$ has not yet been reported.


The effective conduction band mass parameter in In$_2$O$_3$ has been widely studied. Reported effective mass values range from 0.13-0.55~$m_0$ ($m_0$ is the free electron mass),\cite{Feneberg2016,DEWIT1978,Preissler2013,Weiher1962} with rather large spread among results from experimental and theoretical investigations. Experimental results were reported from combined electrical Hall effect and Seebeck effect measurements,\cite{DEWIT1978,Preissler2013,Weiher1962} combined electrical Hall effect and infrared ellipsometry measurements,\cite{Feneberg2016} and angle resolved photoemission measurements.\cite{Scherer2012,Zhang_2011} Many reports of conduction band mass parameters exist which were obtained from first principle calculations in various approximations.\cite{Peelaers2019,Karazhanov2007,Fuchs2008,Odaka_1997,BREWER2004285} Perhaps the most accurate theory result was provided by Fuchs and Bechstedt\cite{Fuchs2008} using first principles density functional theory (DFT) calculations with nonlocal potential resulting from a HSE03 hybrid functional and quasi-particle correction. For the bcc polymorph, the effective electron mass at the $\Gamma$ point is almost direction independent and amounts to 0.22~$m_e$ using the HSE03 functional.\cite{Fuchs2008} The conduction band is nonparabolic, with smaller nonparabolicity than in InN but remains non-negligible.\cite{HofmannJEM_2008} Fuchs and Bechstedt gave an approximation for the conduction band
\begin{equation}
    E_C\left(\mathbf{k}\right)=\frac{1}{2}\left(E_g+\sqrt{E_g^2+4\left(\frac{m_e}{m^{\star}}-1\right)E_{\mathbf{k}}E_g} \right),\label{eq:kp}
\end{equation}
\noindent where $E_{\mathbf{k}}=(\hbar^2/2m_e)\mathbf{k}^2$ and the direct bandgap energy $E_g$ is obtained from $\mathbf{k} \mathbf{p}$ calculations. The nonparabolicity of the effective mass, i.e., its $\mathbf{k}$ dependence, can be determined by matching calculated band structure data against Eq.~\ref{eq:kp}, which permits one to obtain $m^{\star}(\mathbf{k})$ when considering that the inverse of the second derivative of the left side in Eq.~\ref{eq:kp} is related to the $\mathbf{k}$-dependent effective mass parameter
\begin{equation}
    \frac{m^{\star}(\mathbf{k})}{m_e}=\left( 1+\frac{m_e/m^{\star}-1}{1+4(m_e/m^{\star}-1)E_{\mathbf{k}}E_g}\right)^{-1},\label{eq:m(k)}
\end{equation}
\noindent where $m^{\star}$ denotes the effective mass at $\Gamma$.\cite{Fuchs2008} According to Fuchs and Bechstedt\cite{Fuchs2008} the effective mass raises to approximately 0.3~$m_e$ for free electron densities exceeding 10$^{20}$~cm$^{-3}$ in agreement with experimental investigations.  Feneberg~\textit{et al.} used a method combining infrared ellipsometry and electrical Hall effect measurements, introduced by Kasic~\textit{et al.}\cite{KasicPRB2000GaN} for GaN thin films, and reported effective mass parameters for $n$ type In$_2$O$_3$ bulk samples and epitaxial thin films with free electron densities ranging from $5\times 10^{18} \dots 10^{21}$cm$^{-3}$.\cite{Feneberg2016} A strong variation of the effective mass parameter was observed with carrier density, from 0.18~$m_e$ for the lowest detectable carrier density to approximately 0.44~$m_e$ at $10^{21}$cm$^{-3}$, and explained by near $\Gamma$ point conduction band nonparabolicity. A similar conduction band nonparabolicity was observed previously for wurtzite structure InN, where the effective mass varies from 0.05~$m_e$ for perpendicular polarization and 0.037~$m_e$ for parallel polarization at the band bottom to 0.15~$m_e$ for both polarization directions when approaching free electron densities of 10$^{20}$cm$^{-3}$.\cite{HofmannJEM_2008} A simplified approximation for the free electron density $n$ dependence of the conduction band mass $m^{\star}$ was reported by Feneberg~\textit{et al.}
\begin{equation}\label{eq:mvsnFeneberg}
    m^{\star}(n)=\sqrt{m^{\star 2}(0)+2C\hbar^2m^{\star}(0)(3\pi^2n)^{\frac{2}{3}}},
\end{equation}
and matching experimental data resulted in the zero density effective mass parameter $m^{\star}(0)=(0.18\pm0.02)$~m$_0$ and nonparabolocity parameter $C=0.5\pm0.02$~eV$^{-1}$. The zero density effective mass value extrapolated using Eq.~\ref{eq:mvsnFeneberg} is in very good agreement with the $\Gamma$ point effective mass from HSE03 results in Fuchs and Bechstedt. 


The electrical mobility decreases with increasing carrier concentration due to impurity scattering and electron-electron interactions. In$_2$O$_3$ single crystals grown from melt have been studied systematically using different heat treatments at temperatures from 200$^{\circ}$C to 1400$^{\circ}$C.\cite{GalazkaCEC2013} Annealing under non-reducing conditions lead to free electron densities in the 10$^{17}$~cm$^{-3}$ range. Typical electron mobility parameters were reported between 140-180~cm$^2$/(Vs). Such annealed samples revealed transmittance spectra with sharp absorption edge at 440~nm~wavelength with high transparency in the visible range.\cite{GalazkaCEC2013} A new crystal growth technique, “Levitation-Assisted Self-Seeding Crystal Growth Method" with subsequent annealing in O$_2$-containing atmosphere resulted in improved electron mobility of 190~cm$^2$/(Vs) in the low 10$^{17}$~cm$^{-3}$ free electron density range.\cite{GALAZKA2014} Comparison between mobility obtained from electrical Hall effect measurements and broadening parameter showed very good agreement between free electron mobility determined by infrared ellipsometry and electrical Hall effect measurements. It is noted that in this comparison performed by Feneberg~\textit{et al.},\cite{Feneberg2016} electrical mobility is obtained by the same measurement which determines the electron density, and which in turn was used there to determine the effective mass. 


In$_2$O$_3$ is known to develop a surface charge accumulation similarly to indium nitride.\cite{Darakchieva3065030,Nagata_2016,Nagata_2019} It has been posed that this is caused by oxygen vacancies.\cite{Walsh_2011} Additionally, Scherer \emph{et al.} showed that bulk In$_2$O$_3$ crystals cleaved in ultra-high vacuum did not show any electron accumulation layer.\cite{Scherer_2016}
While this effect is necessary for applications in gas sensing, it hinders performance of transistor and diode technologies and must be controlled.


An early vibrational study by White and Keramidas of infrared and Raman spectra from powder In$_2$O$_3$ revealed limited sets of mode parameters. Only 11 out of the 16 transverse optical (TO) mode frequencies and none of the longitudinal optical (LO) mode frequencies were reported.\cite{WHITE1972}  Hamberg and Granqvist, in a large study of sets of indium oxide thin films obtained by evaporation, reported on similar subsets of data.\cite{hamberg_granqvist_1986} Analysis of reflectance measured on pellet samples by Sobatta~\emph{et al.} added the LO mode parameters for the 11 TO mode parameters identified earlier.\cite{Sobotta1990} More recently, thin-film samples have been studied and a complete set of Raman modes have been described.\cite{Garcia-Domene2012, Kranert2014} Infrared ellipsometry was used by Feneberg~\emph{et al.} on a large set of samples in the spectral range above 300~cm$^{-1}$. Highly accurate TO mode frequencies for a subset of 8 modes from the 11 modes already characterized by Sobatta~\emph{et al.} and White and Keramidas were reported. The LO mode parameters were not determined and the focus was directed to determine Drude model parameters for the differently doped samples in order to determine the density dependence of the effective mass parameters. 


Phonon mode parameters obtained from first principles calculations are not available in the literature. 


Hamberg and Granqvist determined the static dielectric constant ($\varepsilon_{\mathrm{DC}} \approx 8.9 - 9.5$) and the high frequency dielectric constant ($\varepsilon_{\infty}=4$) from analysis of broadband reflection and transmission data obtained from polycrystalline thin films deposited by e-beam evaporation.\cite{hamberg_granqvist_1986} No other experimental results for the static dielectric constant appears to be available. Typical capacitance measurement configurations for radio frequency determination of the static dielectric constant require highly insulating material which has not been grown yet. Therefore, capacitance measurements, e.g., as reported for insulating $\beta$-Ga$_2$O$_3$ are not available for In$_2$O$_3$.\cite{Fiedler_2019} Zhang~\textit{et al.} obtained $\varepsilon_{\mathrm{DC}}=9.05$ from DFT calculations.\cite{Zhang_2011} Walsh~\textit{et al.} determined $\varepsilon_{\mathrm{DC}}=9.0$ using interatomic potential calculations.\cite{Walsh2009} The high frequency dielectric constant is mostly agreed upon both by experiment (e.g., Feneberg~\textit{et al.}: $4.08\pm0.02$)\cite{Feneberg2016} and theory (e.g., Fuchs and Bechstedt: 4.0).\cite{Fuchs2008} With the Lyddane-Sachs-Teller relationship,\cite{Lyddane41}
\begin{equation}
    \varepsilon_{\mathrm{DC}}=\varepsilon_{\infty} \prod_{l=1}^{N} \frac{\omega_{\mathrm{LO},l}^{2}}{\omega_{\mathrm{TO},l}^{2}},
    \label{eq:LST}
\end{equation}
\noindent and knowledge of all infrared active transverse (TO; $\omega_{\mathrm{TO}}$) and longitudinal optical (LO; $\omega_{\mathrm{TO}}$) phonon modes and accurate value of $\varepsilon_{\infty}$, $\varepsilon_{\mathrm{DC}}$ can be determined. However, as discussed above only 11 out of the $N=16$ phonon mode pairs, $\omega_{\mathrm{TO},l}$ and $\omega_{\mathrm{LO},l}$, are known so far, while for $l=12, \dots ,16$ neither experimental nor theoretical results seem to exist.

In this work, we report a combined terahertz, far-infrared, and infrared spectroscopic ellipsometry, optical Hall effect, and density functional theory investigation. We perform and analyze measurements on a high quality, unintentionally doped, marginally electrically conductive single crystal of cubic In$_2$O$_3$. We obtain and report the static dielectric constant, the optical phonons, and the free charge carrier properties. We observe and determine all remaining TO and LO phonon mode parameters from our ellipsometry measurements. We also report the complete set of zone center optical phonon modes from our theoretical investigation. We report the static dielectric constant from the full set of optical phonon modes and the Lyddane-Sachs-Teller relation. We further demonstrate excellent agreement with our LST extrapolation by comparing model-calculated with measured THz ellipsometry measurements. Our optical Hall effect measurements reveal density, $n$, mobility, $\mu$, and effective mass parameters which determine an independently measured experimental value closest to the zero density effective mass parameter, $m^{\star}(0)$. We compare our results where available with previous reports, and we compare and discuss our optical Hall effect with electrical Hall effect results.

\section{Theory}
\subsection{Crystal Structure and Symmetry}

\begin{figure}
\centering
\includegraphics[width=.8\linewidth]{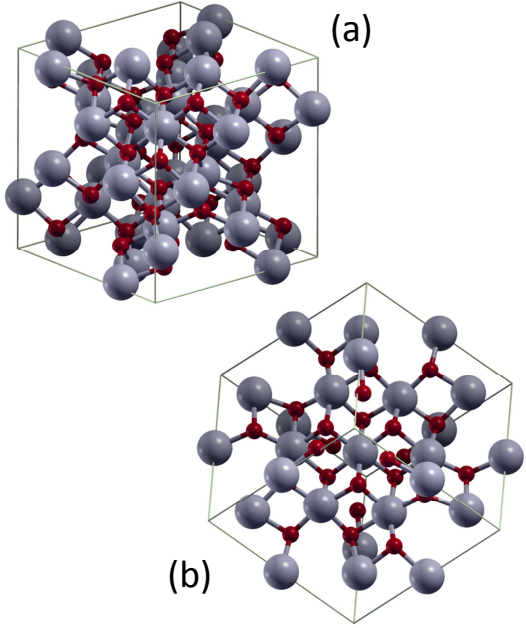}
\caption{\label{fig:Cells} (a) Conventional unit cell of In$_2$O$_3$ and the corresponding (b) primitive cell described in the text. Large (gray) spheres represent indium atoms and the smaller (red) spheres represent oxygen atoms.}
\end{figure}
In$_2$O$_3$ crystallizes in the cubic bixbyte structure (space group No. 206, $Ia{\bar3}$). The primitive cell contains 40 atoms, which results in the total of 117 optical phonon modes, which at the Brillouin-zone center belong to the irreducible representation:
\begin{equation*}
    \Gamma_{\mathrm{opt}} = 4A_{\mathrm{g}} + 5A_{\mathrm{u}} + 4E_{\mathrm{g}} + 5E_{\mathrm{u}} + 14T_{\mathrm{g}} + 16T_{\mathrm{u}}.
\end{equation*}
$E$ and $T$ modes are double- and triple-degenerate, respectively. Modes $A_{\mathrm{g}}$, $E_{\mathrm{g}}$, and 
$T_{\mathrm{g}}$ (g - \emph{gerade}) are Raman-active; modes $T_{\mathrm{u}}$ are IR-active (and the focus of the present study); and modes $A_{\mathrm{u}}$ and $E_{\mathrm{u}}$ (u - \emph{ungerade}) are silent. 

\subsection{Density Functional Theory}
DFT calculations were performed by a plane wave DFT code Quantum ESPRESSO (QE).\cite{[{Quantum ESPRESSO is available from http://www.qu\-an\-tum-es\-pres\-so.org. See also: }]GiannozziJPCM2009QE} We used the local density approximation exchange-correlation functional of Perdew and Wang (PW)\cite{Perdew1992,Perdew2018} and optimized norm-conserving Vanderbilt (ONCV) scalar-relativistic pseudopotentials.\cite{vanSetten2018} As the starting point, we used structural parameters from the Materials Project.\cite{Jain2013, mp22598} [doi: 10.17188/1198812] The initial structure was first relaxed to force levels less than 10$^{-6}$ Ry Bohr$^{-1}$. A regular shifted $2\times2\times2$ Monkhorst-Pack grid was used for sampling of the Brillouin zone.\cite{Monkhorst1976} A convergence threshold of $1\times10^{-12}$ Ry was used to reach self-consistency with the electronic wavefunction cutoff of 120 Ry. The relaxed cell was used for subsequent phonon calculations. The phonon frequencies, Born effective charges, and IR intensities were computed at the $\Gamma$-point of the Brillouin zone using density functional perturbation theory,~\cite{BaroniRMP2001DFTPhonons} as implemented in the Quantum ESPRESSO package, with the convergence threshold for self-consistency of $1\times10^{-18}$ Ry. The parameters of the TO modes were obtained from the dynamical matrix computed at the $\Gamma$-point. The parameters of the LO modes were obtained by setting a small displacement from the $\Gamma$-point in order to include the long-range Coulomb interactions of Born effective charges in the dynamical matrix (the so called non-analytical terms). Mode displacement patterns were rendered using XCrysDen~\cite{[][{. Code available from http://www.xcrysden.org.}]Kokalj1999} running under Silicon Graphics Irix 6.5.

\subsection{Spectroscopic Ellipsometry}
Spectroscopic ellipsometry is a non-invasive and non-destructive measurement technique which measures the change in polarization of light ($\tilde{\rho}$) caused by interaction with the sample. In reflection ellipsometry this change is described by
\begin{equation}
    \tilde{\rho} = \frac{\tilde{r}_{p}}{\tilde{r}_{s}} = \tan(\Psi)e^{i\Delta},
\end{equation}
where $\tilde{r}_{p}$ and $\tilde{r}_{s}$ are the Fresnel reflection coefficients for the light polarized parallel (\emph{p}) to and perpendicular (\emph{s}) to the plane of incidence. The rotation of the light's polarization state about the axis of propagation is defined as $\Psi$ and the relative phase shift between the parallel and perpendicular components is defined as $\Delta$. Hence, for each wavenumber, angle of incidence, and sample position a ($\Psi$, $\Delta$) pair is measured.

\subsection{Four-Parameter Semi-Quantum Model}
To fit for the parameters describing each phonon mode found in the dielectric function, the four-parameter semi-quantum (FPSQ) model first described by Gervais and Periou can be used.\cite{Gervais74,SchubertIRSEBook_2004} This model allows one to fit directly for TO and LO mode frequencies, $\omega_{\mathrm{TO,}l}$ and $\omega_{\mathrm{LO,}l}$, and the determination of their independent broadening parameters, $\gamma_{\mathrm{TO},l}$ and $\gamma_{\mathrm{LO},l}$, respectively
\begin{equation}\label{eq:4PSQmodel}
    \varepsilon = \varepsilon_{\infty} \prod_{l=1}^{N} \frac{\omega^{2}_{\mathrm{LO,}l}-\omega^{2}-i\omega\gamma_{\mathrm{LO,}l}}{\omega^{2}_{\mathrm{TO,}l}-\omega^{2}-i\omega\gamma_{\mathrm{TO,}l}},
\end{equation}
\noindent and $\varepsilon_{\infty}$ is the high-frequency dielectric constant.\cite{SchubertIRSEBook_2004} The product runs over all $N=16$ phonons (TO-LO pairs) with $T_{\mathrm{u}}$ symmetry. The Lyddane-Sachs-Teller (LST) relation\cite{Lyddane41} in Eq.~\ref{eq:LST} then permits calculation of the static dielectric constant, using all other parameters in Eq.~\ref{eq:4PSQmodel} determined from analysis of the wavenumber-by-wavenumber obtained $\varepsilon$.

\subsection{Drude Free Charge Carrier Model}
The FPSQ model is further modified to account for free charge carrier contributions via the addition of a Drude term:\cite{SchubertIRSEBook_2004}
\begin{equation}
    \varepsilon_{\mathrm{FC}} = -\frac{\omega_{\mathrm{p}}^2}{\omega(\omega + i\gamma_{\mathrm{p}})},\label{eq:Drude}
\end{equation}
\noindent where $\omega_{\mathrm{p}}$ is the plasma frequency
\begin{equation}
    \omega_{\mathrm{p}}^{2} = \frac{-\mathrm{e}^2n}{\varepsilon_{\infty}\varepsilon_{0}m^{\star}},
\end{equation}
\noindent and where $n$, $m^{\star}$, and $\gamma_{\mathrm{p}}$ describe the free charge carrier volume density, the effective mass, and the plasma broadening parameter, respectively. From the plasma broadening parameter, the optical mobility parameter, $\mu$, can be found via:
\begin{equation}
    \gamma_{\mathrm{p}} = \frac{\mathrm{e}}{m^{\star}\mu}.
\end{equation}

\subsection{Optical Hall Effect}
The optical Hall effect experiment and analysis procedure is summarized by Schubert~\emph{et al.}\cite{SchubertJOSAA2016OHE} Derived from the equation of motion, the dielectric function component for free charge carriers under the effect of an external magnetic field can be written as follows:
\begin{equation}
    \varepsilon(\omega) = \omega_{\mathrm{p}}^{2}\left[-\omega^2I - i\omega\gamma + i\omega\begin{pmatrix}
  0 & b_3 & -b_2\\ 
  -b_3 & 0 & b_1\\
  b_2 & -b_1 & 0
\end{pmatrix}\omega_{\mathrm{c}}\right]^{-1}.
\end{equation}
Here, $\omega_{\mathrm{c}}$ is the cyclotron frequency
\begin{equation}
    \omega_{\mathrm{c}} = \frac{\mathrm{q}B}{m^{\star}}.
\end{equation}
\noindent Note that the signature of $\omega_{\mathrm{c}}$ depends on the type of the free charge carriers and hence, the signs of the optical Hall effect data reveal $n$ or $p$ type conductivity.\cite{SchubertJOSAA2016OHE} The external magnetic field is given in the ellipsometer coordinate system (\emph{x},\emph{y},\emph{z}) as \(\mathbf{B} = B(b_1,b_2,b_3) \) where \( B = |\mathbf{B}| \). The optical Hall effect experiment gives access to the cyclotron frequency, therefore, if plasma frequency and broadening parameters are known from zero-field ellipsometry measurements, $N_{s}$, $m^{\star}$, and $\mu$ can all be determined independently from each other, and without use of an additional separate experiment, e.g., an electrical Hall effect measurement. One further has the opportunity to compare density and mobility parameters obtained from optical and electrical techniques. It is noteworthy to point out the very different mechanisms of transport which are tested in an optical and in an electrical measurement. For a traditional electrical Hall configuration, for example, transport and scattering phenomena are collected over macroscopic distances across the sample; while for optical measurements, the lateral displacement of the free charge carriers is on the order of nanometers. In an ideal crystal without defects, both techniques will determine very similar if not the same results. 

In these optical Hall effect experiments, the ellipsometry data are recorded using the Mueller matrix convention as opposed to the $\Psi$ and $\Delta$ convention for the zero magnetic field measurements. The Mueller matrix formalism allows accurate capture of the small optical birefringence induced by the optical Hall effect which are then observable in the off-diagonal Mueller matrix elements. The Mueller matrix is used to describe the sample interaction with incoming light described by a Stokes vector as follows:
\begin{equation}
\left( {{\begin{array}{*{20}c}
 {S_{0} } \hfill \\ {S_{1} } \hfill \\  {S_{2} } \hfill \\  {S_{3} } \hfill \\
\end{array} }} \right)_{\mathrm{output}} =
\left( {{\begin{array}{*{20}c}
 {M_{11} } \hfill & {M_{12} } \hfill \ {M_{13} } \hfill & {M_{14} } \hfill \\
 {M_{21} } \hfill & {M_{22} } \hfill \ {M_{23} } \hfill & {M_{24} } \hfill \\
 {M_{31} } \hfill & {M_{32} } \hfill \ {M_{33} } \hfill & {M_{34} } \hfill \\
 {M_{41} } \hfill & {M_{42} } \hfill \ {M_{43} } \hfill & {M_{44} } \hfill \\
\end{array} }} \right)
\left( {{\begin{array}{*{20}c}
 {S_{0} } \hfill \\ {S_{1} } \hfill \\  {S_{2} } \hfill \\  {S_{3} } \hfill \\
\end{array} }} \right)_{\mathrm{input}}.
\end{equation}
\noindent Here, the elements of the Stokes vector are defined as $S_{0}=I_{p}+I_{s}$, $S_{1}=I_{p} - I_{s}$, $S_{2}=I_{45}-I_{ -45}$, $S_{3}=I_{\sigma + }-I_{\sigma - }$. In this definition, $I_{p}$, $I_{s}$, $ I_{45}$, $I_{-45}$, $I_{\sigma + }$, and $I_{\sigma - }$ are the intensities for the $p$-, $s$-, +45$^{\circ}$, -45$^{\circ}$, right handed, and left handed circularly polarized light components, respectively.~\cite{Fujiwara_2007}

\section{Experiment}

\subsection{Crystal Growth}
The investigated crystal sample was prepared from a bulk crystal grown from the melt by a novel technique "Levitation-Assisted Self-Seeding Crystal Growth Method," as described in detail elsewhere.\cite{GALAZKA2013} The free electron concentration was determined from electrical Hall measurements to be $2.65\times 10^{-17}$~cm$^{-3}$, and the Hall electrical mobility is 158~cm$^{2}$/(Vs) (resistivity  0.15~$\Omega$cm) The crystal with surface area of $10\times10$~mm$^2$ and thickness of 500~$\mu$m with (111) surface
orientation was annealed in air at temperatures of 1000$^{\circ}$C for 40~h. Processes for annealing and control of electron density have been reported previously.\cite{GALAZKA2013} The full width at half maximum (FWHM) of the rocking curves of (222) reflexes was $\approx$30 arc sec. The investigated crystal samples were epi-ready polished. The sample was then investigated by ellipsometry and optical Hall effect without further treatment. 

\subsection{Far-infrared and Infrared Ellipsometry}The spectra were measured at room temperature in ambient conditions on two different ellipsometer instruments. The infrared spectral range (650 - 3000~cm$^{-1}$) was measured on a commercial variable angle of incidence spectroscopic ellipsometer (IR-VASE Mark-II; J.A. Woollam Co., Inc.). The far-infrared spectral range (50 - 650~cm$^{-1}$) was measured on an in-house built FIR-VASE instrument.\cite{KuehneRSI_2014} Measurements were performed at $\Phi_{a}$ = 50$^{\circ}$, 60$^{\circ}$, and 70$^{\circ}$ angles of incidence. Multiple azimuthal rotations were not necessary, though they were performed to ensure the cubic symmetry of the optical response. Only one azimuthal rotation is included in this analysis since all rotational data are in excellent agreement with each other. Exemplary experimental data in the FIR range is shown in Fig.~\ref{fig:MM33}. 

\subsection{Optical Hall Effect Experiment}
The same sample was then mounted in our in-house optical Hall effect setup as described by K\"{u}hne~\textit{et al}.\cite{KuehneRSI_2014} Due to the window setup of the superconducting magnet, the angle of incidence is limited to only $\Phi_{a}$= 45$^{\circ}$. The optical Hall effect was measured in the far-infrared and in the infrared range. Measurements were taken at zero field, at $B = +7$~T, and at $B = -7$~T with the sample maintained at room temperature throughout the measurements. The magnetic field was oriented parallel to the incoming beam with the magnetic field strength \(B_{\perp} = B/\sqrt{2} \) perpendicular to the sample surface for all measurements.
\subsection{Terahertz Ellipsometry}
To further investigate the dielectric constants and free charge carrier contributions at quasi-static conditions, ellipsometry was measured in the THz spectrum from 670-900~GHz. The THz ellipsometer is an in-house built instrument operating in the rotating analyzer configuration allowing the measurement of the upper left $3~\times~3$ Mueller matrix elements. Further information on the instrument can be found in K\"{u}hne \textit{et al.}\cite{KuehneRSI_2014} Data from this instrument is shown in Fig.~\ref{fig:THz}.

\section{Results}
\subsection{Density Functional Theory Calculations}

\begin{table}\centering 
\caption{\label{Table:DFT}Parameters for infrared and far-infrared active  phonon modes obtained from DFT calculations in this work. Frequencies, $\omega_{\mathrm{TO}}$ and $\omega_{\mathrm{LO}}$, and intensities, $A^2_{\mathrm{TO}}$ and $A^2_{\mathrm{LO}}$, respectively, of TO and LO modes are presented for all $T_{\mathrm{u}}$ modes. Values for $\varepsilon_{\mathrm{\infty}}$ and $\varepsilon_{\mathrm{DC}}$ are determined as 4.735, and 10.74, respectively, using the LST relationship.}
\begin{ruledtabular}
\begin{tabular}{{l}{c}{c}{c}{c}}  
Mode &  $\omega_{\mathrm{TO}}$  & $\omega_{\mathrm{LO}}$& $A^2_{\mathrm{TO}}$ & $A^2_{\mathrm{LO}}$\\
& (cm$^{-1}$)  &  (cm$^{-1}$) & ($\frac{(\mathrm{D}/\mathrm{\AA})^2}{\mathrm{amu}}$) & ($\frac{(\mathrm{D}/\mathrm{\AA})^2}{\mathrm{amu}}$) \\
\hline
T$_{\mathrm{u}}$-1 & 588.61 &611.35 & 6.284 & 94.215  \\
T$_{\mathrm{u}}$-2 & 553.45 & 569.28 & 5.686 & 33.028  \\
T$_{\mathrm{u}}$-3 & 527.83  & 540.15 & 6.272 & 17.549 \\
T$_{\mathrm{u}}$-4 &  460.07 & 503.84 & 0.075 & 46.718 \\
T$_{\mathrm{u}}$-5 &  409.12 & 460.01 & 48.867 & 0.060 \\
T$_{\mathrm{u}}$-6 &  386.87 & 394.19 & 6.593 & 1.829 \\
T$_{\mathrm{u}}$-7 &  363.37 & 383.47 & 75.398 & 0.888 \\
T$_{\mathrm{u}}$-8 &  330.41 & 339.21 & 37.789 & 0.919 \\
T$_{\mathrm{u}}$-9 &  310.89 & 312.26 & 6.310 & 0.141 \\
T$_{\mathrm{u}}$-10 &  260.02 & 260.09 & 0.139 & 0.011 \\
T$_{\mathrm{u}}$-11 &  214.71 & 216.00 & 1.816 & 0.217 \\
T$_{\mathrm{u}}$-12 &  171.00 & 171.29 & 0.297 & 0.042 \\
T$_{\mathrm{u}}$-13 &  152.81 & 152.85 & 0.038 & 0.006 \\
T$_{\mathrm{u}}$-14 &  148.44 & 148.48 & 0.038 & 0.006 \\
T$_{\mathrm{u}}$-15 &  123.27 & 123.32 & 0.031 & 0.005 \\
T$_{\mathrm{u}}$-16 &  100.86 & 100.86 & $<10^{-4}$ & $<10^{-4}$ \\
\end{tabular}
\end{ruledtabular}
\end{table}

\begin{figure}
\centering
\includegraphics[width=.8\linewidth]{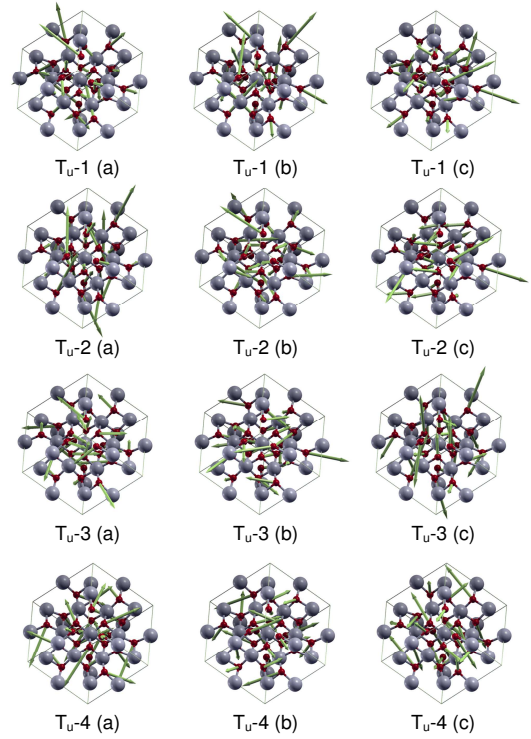}
\caption{\label{fig:Tu14} Rendering of atomic displacement patterns (phonon eigenvectors) for IR-active TO phonon modes T$_{\mathrm{u}}$-1 to T$_{\mathrm{u}}$-4. Note that LO mode displacements, except for different magnitude, are equivalent to their corresponding TO pair partner because of the cubic crystal symmetry.}
\end{figure}

\begin{figure}
\centering
\includegraphics[width=.8\linewidth]{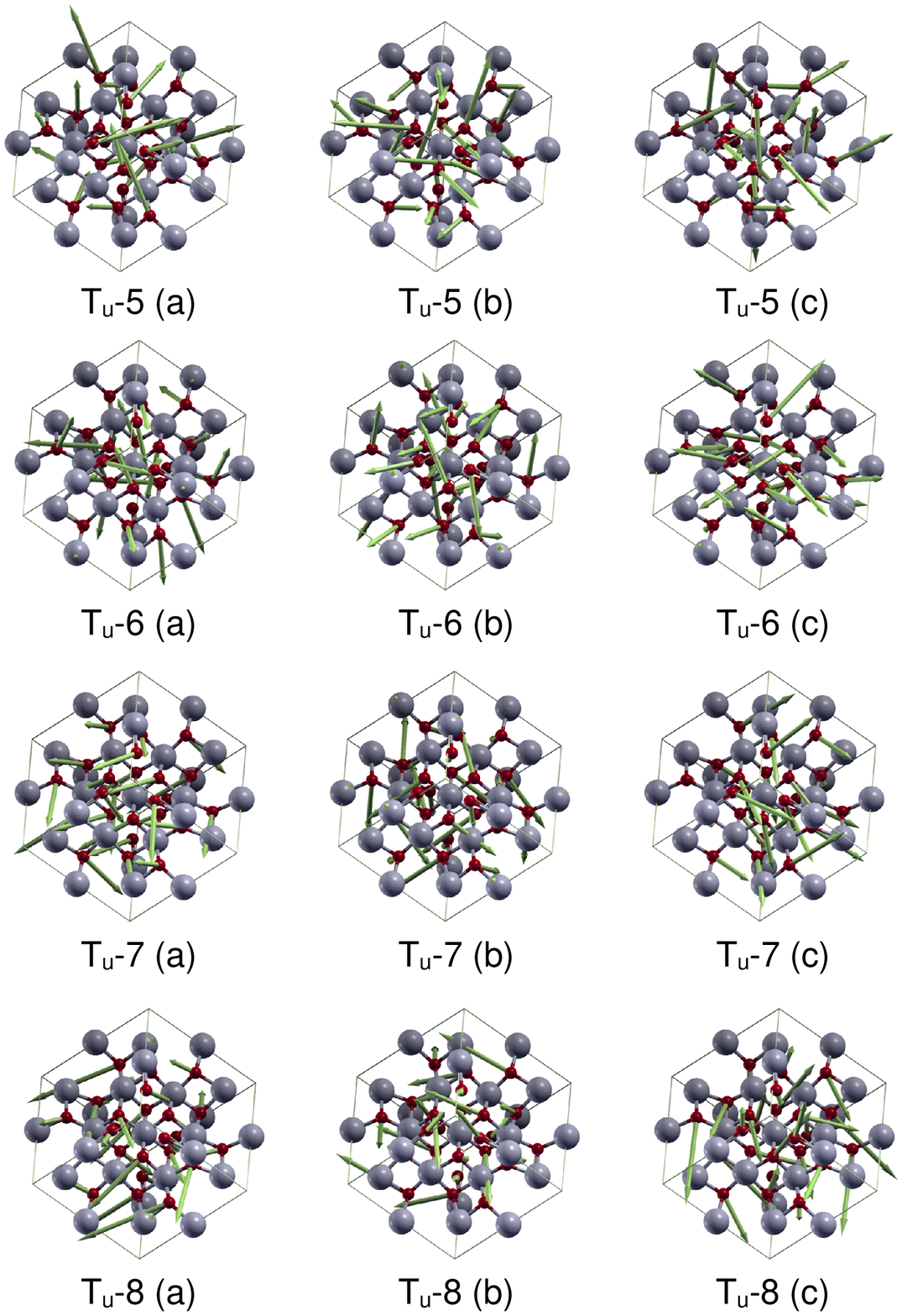}
\caption{\label{fig:Tu58} Same as Fig.~\ref{fig:Tu14} for T$_{\mathrm{u}}$-5 to T$_{\mathrm{u}}$-8.}
\end{figure}

\begin{figure}
\centering
\includegraphics[width=.8\linewidth]{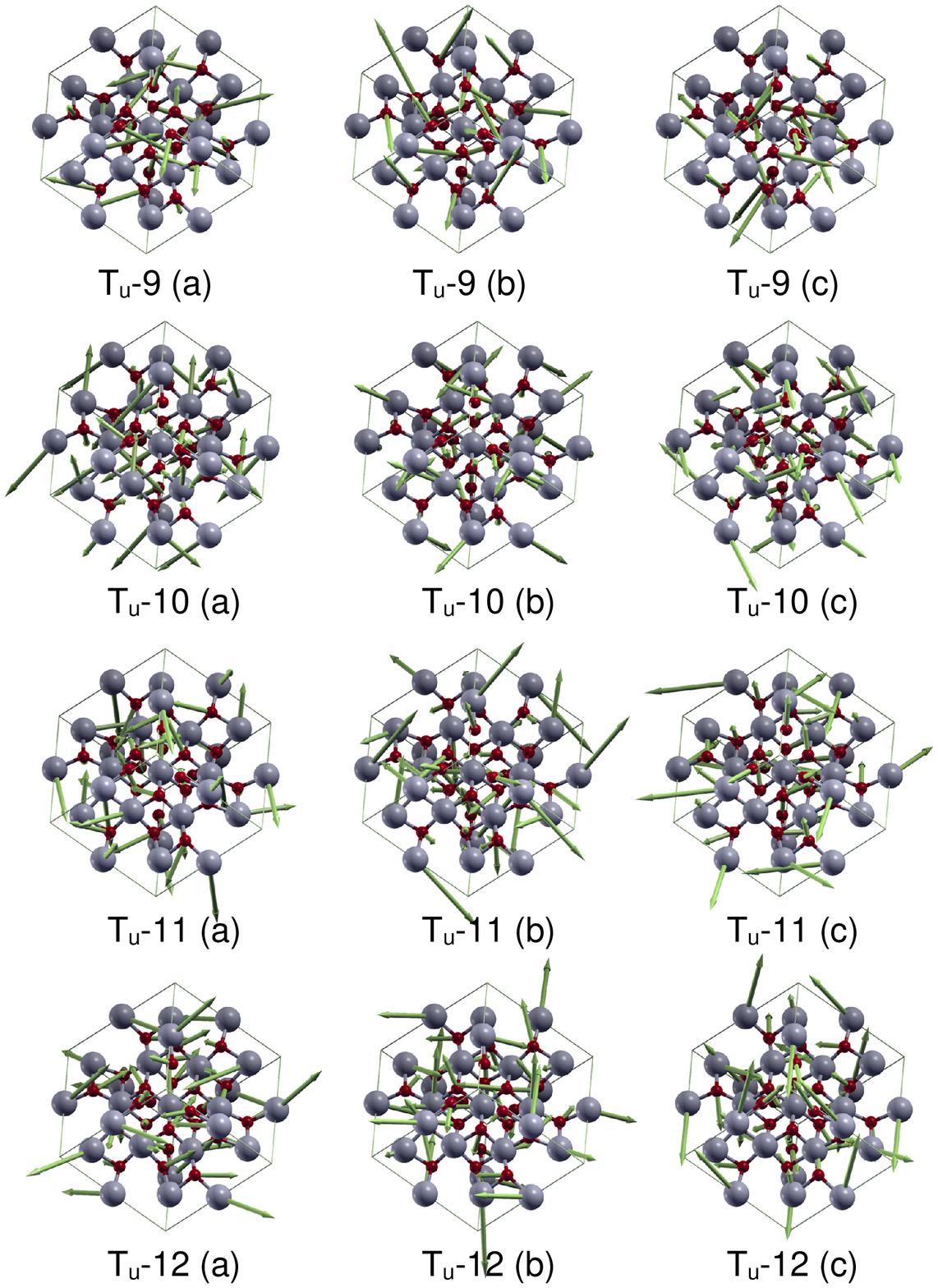}
\caption{\label{fig:Tu912} Same as Fig.~\ref{fig:Tu14} for T$_{\mathrm{u}}$-9 to T$_{\mathrm{u}}$-12.}
\end{figure}

\begin{figure}
\centering
\includegraphics[width=.8\linewidth]{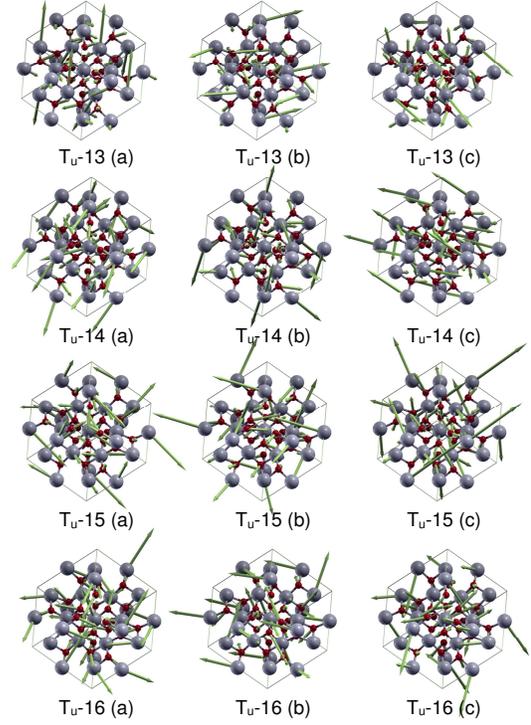}
\caption{\label{fig:Tu1316} Same as Fig.~\ref{fig:Tu14} for T$_{\mathrm{u}}$-13 to T$_{\mathrm{u}}$-16.}
\end{figure}

The parameters of the DFT-calculated TO and LO phonon modes are listed in Table~\ref{Table:DFT} and labelled by $\mathrm{T}_{\mathrm{u}}-i, i = 1, \dots 16$. Renderings of the atomic displacement pattern for the TO modes are shown in Fig.~\ref{fig:Tu14} (modes T$_{\mathrm{u}}$-1 to T$_{\mathrm{u}}$-4), Fig.~\ref{fig:Tu58} (T$_{\mathrm{u}}$-5 to T$_{\mathrm{u}}$-8), Fig.~\ref{fig:Tu912} (T$_{\mathrm{u}}$-9 to T$_{\mathrm{u}}$-12), and Fig.~\ref{fig:Tu1316} (T$_{\mathrm{u}}$-13 to T$_{\mathrm{u}}$-16). Because every T$_{\mathrm{u}}$ mode is triple degenerate, three eigenvectors with labels $a$, $b$, and $c$ are shown for every phonon mode. The corresponding transition dipoles for each such triplet are mutually orthogonal. We note that we observe 16 mode pairs as predicted by group theory. We also note that the intensities of the TO and LO modes are very small for modes 13-16 which may explain why these modes remain undetected thus far.

\subsection{Phonon Mode Analysis from Ellipsometry Data}
\begin{figure}
\centering
\includegraphics[width=0.9  \linewidth]{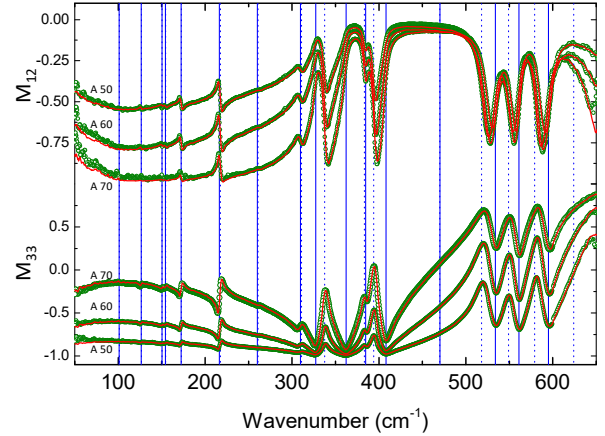}
\caption{\label{fig:MM} Experimental (green open symbols), wavenumber-by-wavenumber best-match model (red solid line) calculated $M_{12}$ and $M_{33}$. Vertical blue solid lines indicate TO modes, and blue dashed lines indicate LO modes identified in this work. Note that the zero-valued off-diagonal elements of the Mueller matrix are excluded as well as elements $M_{21}$ and $M_{22}$ as they show no additional information.}
\end{figure}

\begin{figure}
\centering
\includegraphics[width=  \linewidth]{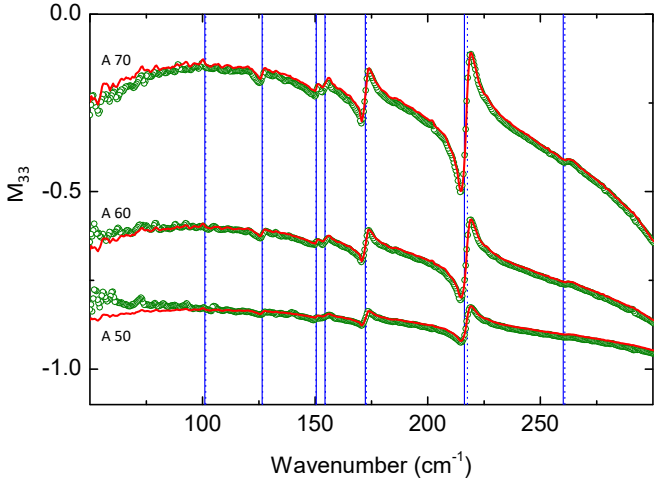}
\caption{\label{fig:MM33} Same as Fig.~\ref{fig:MM} for $M_{33}$ with the far-infrared spectral region expanded emphasizing the spectral range of modes 10-16.}
\end{figure}

To perform this analysis, WVASE32$^{\mathrm{TM}}$ (J.A.~Woollam Co., Inc.) software was utilized. We assume no finite nanometer-scale surface roughness as such roughness does not contribute to the ellipsometric data in the infrared and far-infrared spectral regions.\cite{SchubertIRSEBook_2004} It is worth noting here that if a surface accumulation layer had been present on this sample, it was not observable in optical Hall effect measurements nor in our zero field spectroscopic ellipsometry measurements. Initial analysis is done using a wavenumber-by-wavenumber approach where a model dielectric function is allowed to fit for each wavenumber independently to find a best-match model of the real and imaginary dielectric values. In a subtle modification to this approach, we also added a free charge carrier contribution to the model dielectric function, according to Eq.~\ref{eq:Drude}. This step is performed to reduce the otherwise small increase in the imaginary part of $\varepsilon$ towards the far-infrared spectral range, which is caused by the absorption of the low density free electrons present in our sample. The resulting Drude model parameters will be discussed further below. The resulting dielectric function with added free charge carrier contributions lead to the therefrom calculated Mueller matrix data which excellently match with the experimental data in Fig.~\ref{fig:MM}. Selected experimental and best-model calculated data in the far-infrared spectral region emphasizing small yet distinct signatures of phonon modes unidentified previously in the long wavelength range are shown in Fig.~\ref{fig:MM33}.

\begin{figure}
\centering
\includegraphics[width=\linewidth]{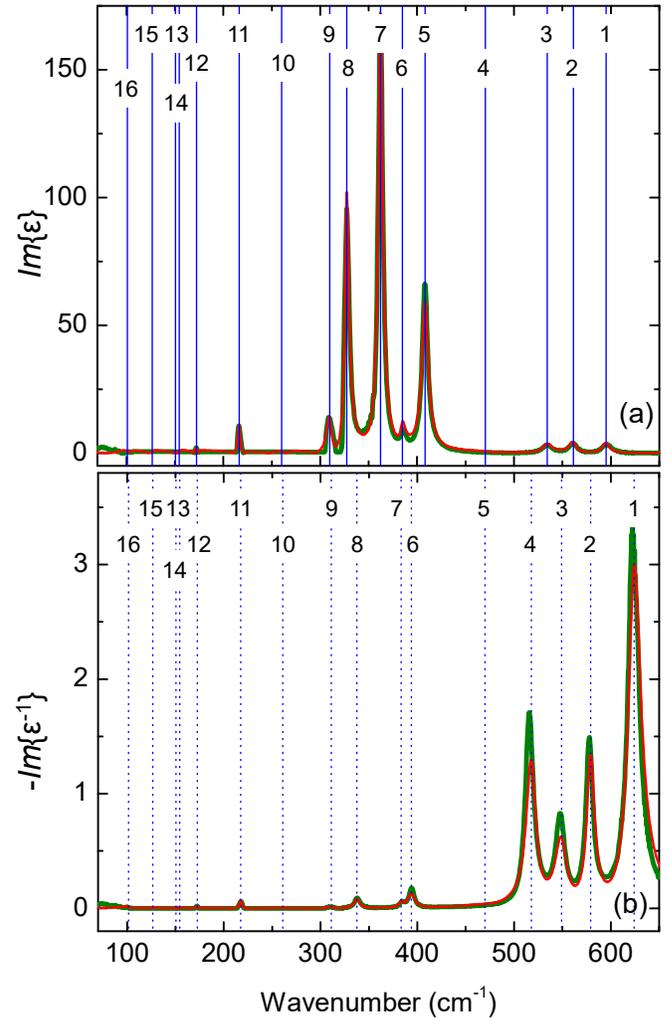}
\caption{\label{fig:Epsilon} Wavenumber-by-wavenumber best-match model calculated (green solid lines) and best-match model dielectric function calculated (red solid line) showing the imaginary (a) and innverse imaginary (b) parts of the dielectric function. Vertical blue solid lines indicate TO modes and blue dotted lines indicate LO modes identified in this work.}
\end{figure}

\begin{figure}
\centering
\includegraphics[width=\linewidth]{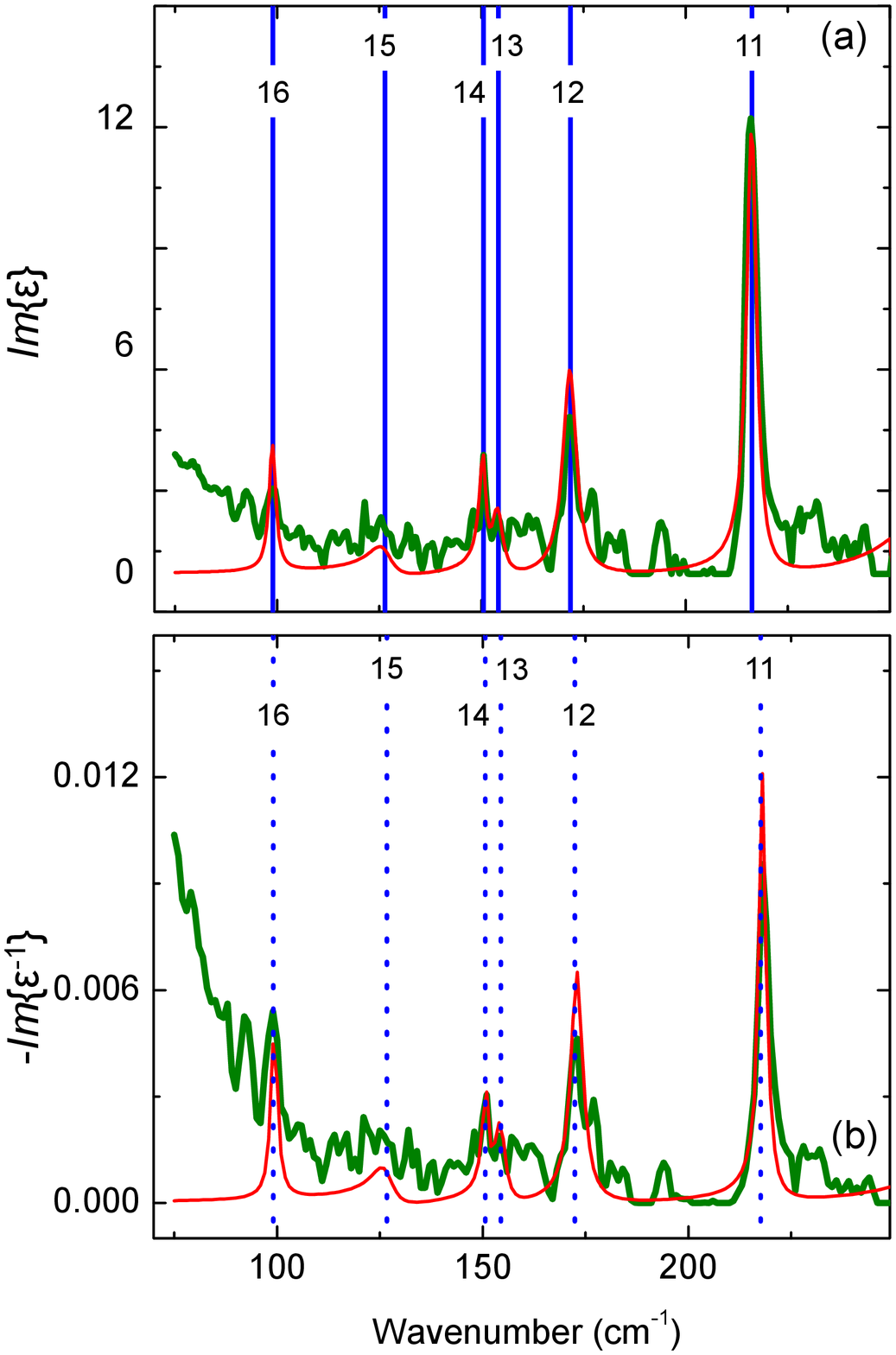}
\caption{\label{fig:FarEpsilon} Same as Fig.~\ref{fig:Epsilon} with the far-infrared spectral region expanded emphasizing the spectral range of modes 11-16.}
\end{figure}

Figure~\ref{fig:Epsilon} summarizes the wavenumber-by-wavenumber best-model parameter $\varepsilon(\omega)$ and Fig.~\ref{fig:FarEpsilon} emphasises the far-infrared spectral region. We were then able to best-model match the dielectric function using the FPSQ model in Eq.~\ref{eq:4PSQmodel}. Implementing sixteen TO-LO pair parameters and $\varepsilon_{\infty}$ we were able to find excellent agreement between the best-match model and the wavenumber-by-wavenumber produced dielectric function, as shown in Fig.~\ref{fig:Epsilon}. Note that we present the imaginary parts of $\varepsilon$ and $\varepsilon^{-1}$ since these immediately reveal frequency and broadening of the TO and LO mode parameters, as one can assess from Eq.~\ref{eq:4PSQmodel}. We note further that small additional features in the far-infrared range, Fig.~\ref{fig:FarEpsilon}, are most likely due to noise increasing rapidly for wavenumbers below 100~cm$^{-1}$ towards the THz range, best seen in Fig.~\ref{fig:MM33}. This noise is due to the loss of detected intensity in the far-infrared probe beam. All best-match model parameters and uncertainties are summarized in Table~\ref{Table:Ellip}. We identify all mode pairs predicted by our DFT calculations and their broadening parameters. We note that the generalized Lowndes condition is fulfilled, where the sum of all LO broadening parameters is larger or equal to the sum of all TO broadening parameters.\cite{KasicPRB2000GaN} While several modes are small in the lower spectral range, they remain detectable and were able to be fit for with limited interference. All modes are indicated in Fig.~\ref{fig:Epsilon}. Values which were range limited are designated in Table~\ref{Table:Ellip} with an asterisk and as a result have very large uncertainty. This can be attributed to the small magnitude of the modes (seen in the DFT analysis as well) and growing noise in the lower spectral range. 

Also shown in Table~\ref{Table:Ellip} are previously reported TO phonon frequencies as found by Feneberg~\textit{et al.}\cite{Feneberg2016} and TO and LO mode frequencies as observed by Sobotta~\textit{et al.}\cite{Sobotta1990} Our values are in excellent agreement with the eight TO modes determined by Feneberg~\textit{et al.}, except for the small mode T$_{\mathrm{u}}-4$, which was not observed there. Also, Feneberg~\textit{et al.} did not determine any LO mode parameters, and modes below ~300~\rzcm remained inaccessible. The phonon modes reported by Sobotta~\textit{et al.} is the most comprehensive data set to date, and is in reasonable agreement with our results, except for the lack of the far-infrared modes. We note that specifically LO modes are inherently difficult to obtain from reflectance data of pellets and low quality samples which may explain some of the discrepancies between data of Sobotta~\textit{et al.} and our results. Modes 12-16 remained unobserved previously and can only be compared against our own DFT calculations. We note the overall very good agreement between our experimental and DFT results.


\begin{figure}
\centering
\includegraphics[width=\linewidth]{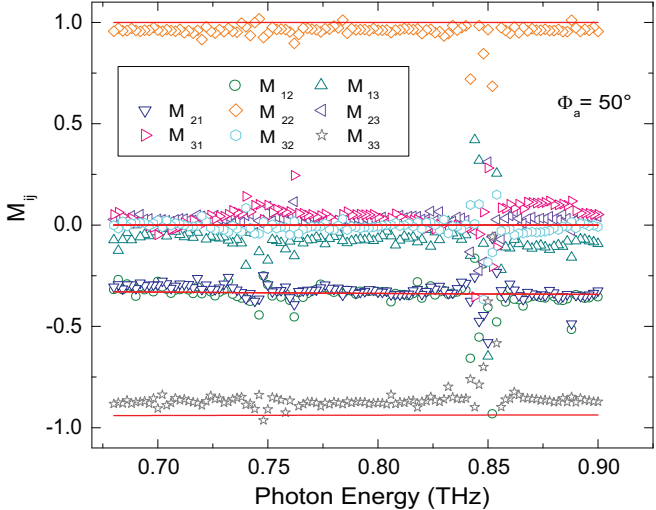}
\caption{\label{fig:THz} Zero-magnetic-field THz Mueller matrix ellipsometry data $M_{ij}$ measured at 50$^{\circ}$ angle of incidence (Symbols: experiment; solid lines: best model calculated data). A very good agreement is observed. Noise at approximately 0.75 and 0.85~THz is due to atmospheric absorption. The best-match model calculated data are obtained using  $\varepsilon_{\mathrm{DC}}=10.55$ and the Drude model parameters $\omega_p$ and $\gamma_p$ parameterized using the result from the optical Hall effect measurement shown in Fig.~\ref{fig:OHE}. No further parameter variation was done to match the THz data spectral range.}
\end{figure}

\begin{table*}\centering 
\caption{\label{Table:Ellip}FPSQ model parameters for TO and LO modes obtained from best-match model dielectric function analysis. $\varepsilon_{\mathrm{\infty}}$ and $\varepsilon_{\mathrm{DC}}$ are determined as 4.05 and 10.55 using the LST relationship. Included are data obtained by Feneberg~\textit{et al.}\cite{Feneberg2016} and Sobottoa~\textit{et al}.\cite{Sobotta1990} observed by infrared ellipsometry and combined far-infrared and infrared reflectance, respectively. All numericals in units of cm$^{-1}$.} 
\begin{ruledtabular}
\begin{tabular}{{l}{c}{c}{c}{c}{c}{c}{c}}
 & \multicolumn{3}{c}{\scriptsize{$\omega_{\mathrm{TO}}$}} & \multicolumn{2}{c}{\scriptsize{$\omega_{\mathrm{LO}}$}} &\multicolumn{1}{l}{} & \multicolumn{1}{l}{} \\\cline{2-4} \cline{5-6} 
\scriptsize{Mode}& \scriptsize{This work} & \scriptsize{Ref.~\onlinecite{Feneberg2016}} &\scriptsize{Ref.~\onlinecite{Sobotta1990}} &\scriptsize{This work}  &\scriptsize{Ref.~\onlinecite{Sobotta1990}}&\scriptsize{$\gamma_{\mathrm{TO}}$}&\scriptsize{$\gamma_{\mathrm{LO}}$}\\
\hline
T$_{\mathrm{u}}$-1  & 594.9 ± 0.7 & 595.5 & 612& 624 ± 0.3 & 625 & 8.83 ± 0.4 & 16 ± 0.5 \\
T$_{\mathrm{u}}$-2  & 561.0 ± 0.9 & 560.5 & 570& 579.1 ± 0.3 & 578& 8.59 ± 0.7  & 10.0 ± 0.8 \\
T$_{\mathrm{u}}$-3  & 534.0 ± 1.3 & 534.1 & 542& 549.1 ± 5.9 & 548& 11.0 ± 0.9 & 14.7 ± 0.9 \\
T$_{\mathrm{u}}$-4  & 470.0 ± 0.1 & -     & 489 & 518.0 ± 0.6 & 513 & 21 ± 0.2     & 12* \\
T$_{\mathrm{u}}$-5 & 408.2 ± 0.2 & 408.0 & 409 & 470.1 ± 0.2 & - & 8.0*  & 23.2 ± 0.1       \\
T$_{\mathrm{u}}$-6  & 384.7 ± 0.1 & 386.2 & 380& 394.0 ± 0.1 & 391& 5.4 ± 0.2  & 8.5 ± 0.1 \\
T$_{\mathrm{u}}$-7  & 362.2 ± 0.2 & 361.7 & 362 & 383.2 ± 0.1 & 381& 5.6 ± 0.3  & 8.5 ± 0.2 \\
T$_{\mathrm{u}}$-8  & 327.3 ± 0.1 & 327.3 & 322& 337.4 ± 0.1 & 352 & 5.3 ± 0.2  & 6.9 ± 0.1  \\
T$_{\mathrm{u}}$-9  & 309.7 ± 1.2 & 307.3 & 301& 310.4 ± 1.1 & 318& 6.5 ± 1.1  & 5.4 ± 0.1  \\
T$_{\mathrm{u}}$-10 & 260.2 ± 5.3 & -      & 245& 260.9 ± 4.7 & 245& 16.7 ± 0.1 & 15.6 ± 0.1      \\
T$_{\mathrm{u}}$-11 & 216.3 ± 0.4 & -      & 205& 217.6 ± 0.2 &224 & 3.0 ± 0.4  & 2.8 ± 0.4 \\
T$_{\mathrm{u}}$-12& 172 ± 0.9 & -     & -& 172.4 ± 0.1 & -& 4.2 ± 0.9  & 4.1 ± 0.2 \\
T$_{\mathrm{u}}$-13 & 154.2 ± 1.1 & -      &- & 154.4 ± 1.1 &- & 3.5 ± 1.2       & 3.5 ± 0.4      \\
T$_{\mathrm{u}}$-14 & 150.5 ± 1.2 & -      &- & 150.6 ± 2.1 & -& 2.1 ± 0.9  & 2.0 ± 0.2 \\
T$_{\mathrm{u}}$-15 & 126.4 ± 1.7 & -      &-& 126.7 ± 0.3  &- & 6.9 ± 2.4        & 6.8 ± 2.4        \\
T$_{\mathrm{u}}$-16 & 99.0 ± 0.2 & -      & -& 99.1 ± 0.2 & -& 2.3 ± 0.8      & 2.3 ± 0.8      
		\end{tabular}
\end{ruledtabular}
\end{table*}

\subsection{Free Electron Parameters}
\begin{figure}
\centering
\includegraphics[width=\linewidth]{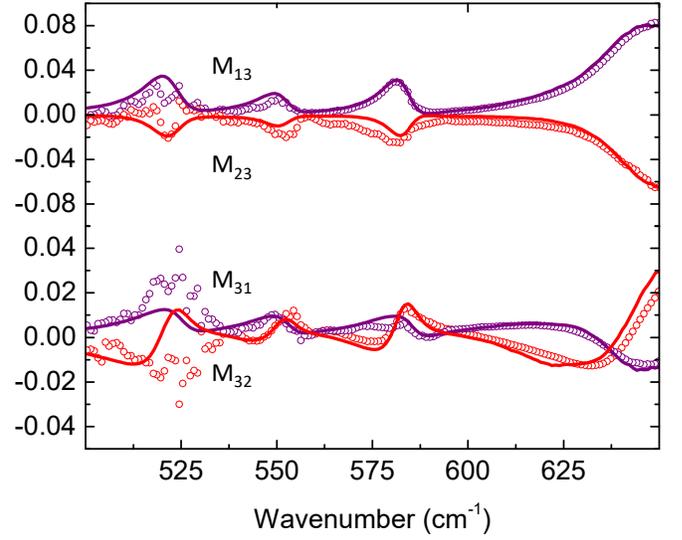}
\caption{\label{fig:OHE} Experimental (symbols) and best-match model calculated (solid lines) optical Hall effect data as difference data between $+7$~T and $-7$~T at $\phi_{a}$= 45$^{\circ}$ angle of incidence. The model calculated lines are obtained using the  wavelength-by-wavelength data for $\varepsilon$ shown in Fig.~\ref{fig:Epsilon} and the Drude model parameters $n$~=~$2.81\times~10^{17}$~cm$^{-3}$, $m^{\star}$~=~(0.208~$\pm$~0.006)~$m_0$, and $\mu$~= ~(112~$\pm$~$3)~\mathrm{cm^{2}/(Vs)}$.}
\end{figure}

\begin{figure}
\centering
\includegraphics[width=\linewidth]{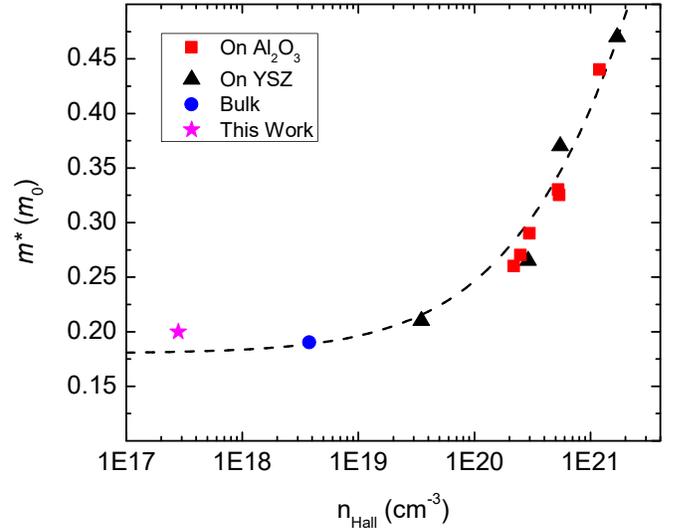}
\caption{\label{fig:Effmass} Data from Ref.~\onlinecite{Feneberg2016} showing the evolution of effective mass against the free charge density for different sample types with the addition of our sample. Here, the nonparabolic band dependence model of the effective mass is shown with the dashed line according to Feneberg~\textit{et al.} in Ref.~\onlinecite{Feneberg2016}.}
\end{figure}

The free electron parameters were characterized via optical Hall effect measurements and compared to electrical Hall effect results. Data are shown in Fig.~\ref{fig:OHE}. We note that for analysis of the optical Hall effect data only one more parameter is needed, $\omega_{\mathrm{c}}$. The other two parameters, $\omega_{\mathrm{p}}$ and $\gamma_{\mathrm{p}}$, are already determined during the phonon mode analysis step by matching the zero field ellipsometry parameters to the best-model dielectric function. As discussed previously,\cite{SchubertJOSAA2016OHE} the optical Hall effect signatures are proportional to $\omega_{\mathrm{c}}$, and hence, combining zero field measurements with optical Hall effect measurements provides the third parameter from experiment. This then leads to determination of $n$, $\mu$ and $m^{\star}$. Here we find $n=(2.81\pm0.01)\times~10^{17}$~cm$^{-3}$ in excellent agreement with the electrical Hall effect density of $n=2.65\times~10^{17}$)~cm$^{-3}$. These values are also in agreement with those reported by Galazka on similar samples in 2014.\cite{GALAZKA2014} Due to this relatively low carrier density, the optical Hall effect data signatures are markedly small. Nonetheless, our best-match model provides a consistent fit. We also obtain the effective mass $m^{\star}$~=~(0.208~$\pm$~0.006)~$m_0$ which is very similar to the zero density value, $m^{\star}(0)=0.18~$m$_0$ reported by Feneberg~\textit{et al.}\cite{Feneberg2016} This value was obtained from matching a nonparabolic band approximation model for the effective mass parameter to a set of samples from which effective mass and free electron density parameters were determined. This data set and best-match function $m^{\star}(n)$ is reproduced in Fig.~\ref{fig:Effmass}, where our result is included. The effective mass measured in our work by optical Hall effect corresponds to one order of magnitude lower electron density than reached by Feneberg~\textit{et al.}, and is thus closer to the zero density value. Within our error bar, our value matches excellently with the zero density effective mass extrapolated previously. We also find the free electron mobility, $\mu =(112 \pm 3)$~cm$^{2}$/(Vs) from optical Hall effect which is slightly lower than the value of 158~cm$^{2}$/(Vs) obtained from electrical Hall effect, which is also within the range reported by Galazka, Uecker, and Fornari on similar oxygen annealed samples $(130-190$~cm$^{2}$/(Vs)).\cite{GALAZKA2014}

\subsection{Static and high frequency dielectric constants}

In order to determine the high frequency dielectric constant, we extended the spectral range for our wavelength-by-wavelength analysis far above the phonon mode range, to 3000~cm$^{-1}$. At frequencies far above the phonon modes yet far enough below the onset of excitonic and band-to-band transitions so that dispersion introduced by the latter is negligible, a good approximation for $\varepsilon_{\infty}$ can be found. Our analysis resulted in 4.05 $\pm$ 0.05. This is in good agreement with our DFT predicted value (4.128) as well as the value previously reported by Feneberg \textit{et al.} (4.08).\cite{Feneberg2016} Using the LST relationship, the static dielectric constant, $\varepsilon_{\mathrm{DC}}$, can then be calculated since we have determined the complete set of phonon modes. With the values listed in Table~\ref{Table:Ellip} we obtain here 10.55 $\pm$ 0.07. This value is in agreement with our DFT value of 10.74. We further conducted ellipsometry measurements at THz frequencies in order to directly evaluate the static dielectric permittivity parameters. While the free carrier concentration is small (the smallest for which effective mass parameters have been reported so far), due to the rather thick sample, our sample is opaque in the THz range.  Hence, THz measurements can only be made in reflection and no substrate interferences can be detected. Such Fabry-Perot interference fringes increase sensitivity to the quasi-static DC permittivity, as demonstrated recently for $\beta$-Ga$_2$O$_3$.\cite{GopalanAPL2020} Figure~\ref{fig:THz} depicts measured and best-model calculated THz Mueller matrix data measured at 50$^{\circ}$ angle of incidence. The best-model calculated data were obtained without further parameter variations only using the static dielectric constant and the Drude model in Eq.~\ref{eq:Drude} contributions with the parameters for $n$, $\mu$, and $m^{\star}$ from optical Hall effect as discussed above. The very good agreement between measured and model-calculated THz Mueller matrix data is indicative of the correctness of the static DC permittivity for In$_2$O$_3$, which has not been determined from experiment at DC or quasi-static frequencies previously.

\section{Conclusions}
By using a combined approach of spectroscopic ellipsometry, DFT calculations, and optical Hall effect measurements we are able to provide a thorough investigation of the electrical and optical phonon mode properties of In$_2$O$_3$. All sixteen TO-LO pairs have been identified and their respective broadening parameters quantified. These values find excellent agreement with the limited sets of phonon mode information previously reported, and we introduce phonons in the far-infrared spectrum not detected thus far. Also our measured phonon modes are in excellent agreement with the results of our DFT calculations. Furthermore, by means of optical Hall effect measurements, we determine the effective electron mass at the lowest yet detected free electron density, which is in excellent agreement with the zero density effective electron mass in In$_2$O$_3$ predicted previously from extrapolation.

\section{Acknowledgments}
This work was supported in part by the National Science Foundation (NSF) under award NSF DMR 1755479, NSF DMR 1808715, by the Nebraska Materials Research Science and Engineering Center award DMR 1420645, in the framework of GraFOx, a Leibniz-Science Campus partially funded by the Leibniz Association - Germany, by Air Force Office of Scientific Research under awards FA9550-18-1-0360 and FA9550-19-S-0003, by the Swedish Research Council VR award No. 2016-00889, the Swedish Foundation for Strategic Research Grant Nos. RIF14-055 and EM16-0024, by the Swedish Governmental Agency for Innovation Systems VINNOVA under the Competence Center Program Grant No. 2016–05190, by the Knut and Alice Wallenbergs Foundation supported grant 'Wide-bandgap semiconductors for next generation quantum components', by the Swedish Government Strategic Research Area in Materials Science on Functional Materials at Link{\"o}ping University, and Faculty Grant SFO Mat LiU No. 2009-00971. Mathias Schubert acknowledges the University of Nebraska Foundation and the J.~A.~Woollam~Foundation for financial support.

The data that support the findings of this study are available from the corresponding author upon reasonable request.
\bibliography{INOlibrary}
\end{document}